\batchmode
\documentclass[english,aps,preprint]{paper}
\usepackage[T1]{fontenc}
\usepackage[latin9]{inputenc}
\usepackage{geometry}
\usepackage{textcomp}
\usepackage{amsmath}
\usepackage{amssymb}
\usepackage{setspace}
\makeatletter

\DeclareRobustCommand{\greektext}{%
  \fontencoding{LGR}\selectfont\def\encodingdefault{LGR}}
\DeclareRobustCommand{\textgreek}[1]{\leavevmode{\greektext #1}}
\DeclareFontEncoding{LGR}{}{}
\DeclareTextSymbol{\~}{LGR}{126}
\newcommand{\lyxmathsym}[1]{\ifmmode\begingroup\def\b@ld{bold}
  \text{\ifx\math@version\b@ld\bfseries\fi#1}\endgroup\else#1\fi}

\makeatother

\usepackage{babel}
\begin{document}

\title{Aspects of the Coarse-Grained-Based Approach to a Low-Relativistic
Fractional Schrödinger Equation}

\author{J. Weberszpil$^{1}$, C.F.L. Godinho$^{2}$, A. Cherman$^{3,4}$
, J. A. Helayël-Neto$^{4}$}

\maketitle
\textit{$^{1}$josewebe@ufrrj.br}

\textit{Universidade Federal Rural do Rio de Janeiro, UFRRJ-IM/DTL}\\
\textit{ Av. Governador Roberto Silveira s/n- Nova Iguaçú, Rio de
Janeiro, Brasil}

\textit{$^{2}$crgodinho@ufrrj.br}

\textit{Grupo de Física Teórica, Departamento de Física, Universidade
Federal }\\
\textit{Rural do Rio de Janeiro, BR 465-07, 23890-971, Seropédica,
RJ, Brasil}

\textit{$^{3}$acherman@pcrj.rj.gov.br}

\textit{Fundação Planetário da Cidade do Rio de Janeiro,}\\
\textit{Rua Vice-Governador Rubens Bernardo, 100, 22451-070, Rio de
Janeiro, Brasil}

\textit{$^{4}$helayel@cbpf.br}

\textit{Centro Brasileiro de Pesquisas Físicas-CBPF-Rua Dr Xavier
Sigaud 150,}\\
\textit{ 22290-180, Rio de Janeiro RJ Brasil. }
\begin{abstract}
The main goal of this paper is to set up the coarse-grained formulation
of a fractional Schrödinger equation that incorporates a higher (spatial)
derivative term which accounts for relativistic effects at a lowest
order. The corresponding continuity equation is worked out and we
also identify the contribution of the relativistic correction the
quantum potential in the coarse-grained treatment. As a consequence,
in the classical regime, we derive the sort of fractional Newtonian
law with the quantum potential included and the fractional conterparts
of the De Broglies's energy and momentum relations.
\end{abstract}

\section{Introduction}

Physicists are presently seeking and trying to understand the connections
between complex systems, nonlocal field theories and other areas of
Physics. This is today an important subject of studies in different
physical and mathematical areas, but the understanding of non-linear
processes connected to these topics has had a considerable boost over
the past 40 years. This deeper comprehension has been inspired by
the discovery and the insight of a new phenomenon, known as dynamical
chaos. The main motivation is that the use of these theories may yield
a much more elegant and effective treatment of problems in particle
and high energy physics, as it has hitherto been carried out with
the help of the local field theories. A particular subclass of non-local
field theories is described with the operators of a fractional nature
and is specified in the framework of fractional calculus (FC). FC
provides us with a set of mathematical tools to generalize the concept
of derivative and integral operators with integer order to their respective
extensions of an arbitrary real order. FC has raised up a great deal
of interest over recent years and has been used as an applied tool
to the the study of fractional dynamics in many fields of physics,
mechanics, engineering and other areas to approach problems connected
with complex systems \cite{Richard Herrmann}. Today, there is a rich
stream of works linking such areas throughout different paths, \cite{Cresus e Everton}.
Non-local theories and memory effects can also be connected to complexity
and admit a treatment in terms of FC. In this context, the non-differentiable
nature of the microscopic dynamics may be connected with time scales
so as to approach questions in the realm of complex systems \cite{Cresus-Jose-Helayel-caos}.

The inclusion of relativistic effects into the Schrödinger equation
intend to the correct computation of the atomic spectrum and in the
area of heavy-ion collisions, relativistic contributions are typically
much larger and, especially for atoms with large nuclear charges Z
, these effects can be quite significant \cite{Q. Su}. Relativistic
contributions have also been considered in the studies of the electron
motion in the operation of free-electron lasers. Various important
characteristics of a quantum system may not be well determined if
those relativistics effects are not completely taken into account
in calculation, e.g in the context of the atomic and molecular structure
and the energy levels, in the describing of excited states and the
fine structure of an hydrogen like atoms and so on.

According to K. G. Dyall \cite{Dyall}, experimental evidences with
atomic structure has shown that calculations which proceed from a
fully relativistic model are longer than the corresponding nonrelativistic
calculations. To ascertain what level of accuracy is required, we
have to make the inclusion of relativistic effects that themselves
makes a significant difference to the results, or whether the error
in including them by perturbation theory is significant. Neglecting
this picture change may lead to serious inaccuracies, e.g., the calculations
for order-of-magnitude estimates of a quantity may not need to consider
relativity at all, whereas calculations of experimental accuracy on
the hydrogen atom must account for relativistic effects in detail.
In making an assessment of whether relativistic effects should be
included, and by what means, it is useful to have an overview of the
effect of relativity on structure, the size of relativistic effects
in the periodic system and some criteria for judging their importance.
Also, following M. Reither and B. Heß \cite{Reither}, this description
are also important for the interpretation of highly accurate experiments
in spectroscopy. As already mentioned, the so-called relativistic
effects begin to play a major role in heavy atoms and their compounds.
This is due to the fact that the relativistic effects on energies
and other physical quantities increase with the fourth power of the
nuclear charge Z.

The relativistic linear Schrödinger equation has been discussed at
the early years of quantum mechanics but was dismissed promptly by
the Klein-Gordon and the Dirac equations. Recently, relativistic versions
of the Schrödinger equation have been considered in the study of relativistic
quark-anti-quark bound states \cite{Nickisch}, and gravitational
collapse of a boson star \cite{Frohlich J and Lenzmann E}.

In the present, we pursue an investigation of the coarse-grained fractional
Schrödinger equation corrected by fourth spatial derivative term which
accounts for a relativistic correction to the kinetic energy term
in the Hamiltonian.

Recently, Guy Jumarie \cite{Jumarie1} proposed an alternative definition
to the Riemann-Liouville derivative. His modified Riemann-Liouville
derivative (MRL) has the advantages of both the standard Riemann-Liouville
and Caputo fractional derivatives: it is defined for arbitrary continuous
(non-differentiable) functions and the fractional derivative of a
constant is equal to zero. The MRL approach seems to give a mathematical
framework for dealing with dynamical systems defined on coarse-grained
spaces and with coarse-grained time and, to this end, to use the fact
that fractional calculus appears to be intimately related to fractal
and self-similar functions. The well-tested definitions for fractional
derivatives, namelly, Riemann-Liouville and Caputo have been frequently
used for several applications. In spite their adequacy, they have
some dangerous pitfalls. For this reason we use here the MRL approach
of fractional derivative. Basic definitions \cite{Jumarie1,Cresus-Jose-Helayel-caos,Jumarie-Lagrang Fract,Livro Jumarie,Jumarie- Acta Sin,Jumarie2}
and detail of the formalism can be found in the cited references and
references therein.

We would like to emphasize that the choice of MRL approach, besides
the points already mentioned, is justified by the fact that the chain
and Leibniz rules acquires a simpler form, which helps a great deal
if changes of coordinates are performed. Moreover, causality seems
to be more easily obeyed in a field-theoretical construction if we
adopt this approach.

In a previous work \cite{Cresus-Jose-Helayel-caos},\textbf{ }we have
argued that the modelling of TeV-physics may demand an approach based
on fractal operators and FC. We claimed that, in the realm of complexity,
non-local theories and memory effects were connected to complexity
and the FC and that the non-differentiable nature of the microscopic
dynamics may be connected with time scales. Using the MRL definition
of fractional derivatives, we have worked out explicit solutions to
a fractional wave equation with suitable initial conditions to carefully
understand the time evolution of classical fields with a fractional
dynamics. First, by considering space-time partial fractional derivatives
of the same order in time and space, a generalized fractional D'Alembertian
is introduced and by means of a transformation of variables to light-cone
coordinates, an explicit analytical solution were obtained. Also,
aspects connected with Lorentz symmetry were analyzed in two different
approaches.

Here, we claim that the use of an approach based on a sequential form
of MRL \cite{Jumarie1} is more appropriate to describe the dynamics
associated with field theory and particle physics in the space of
non-differentiable solution functions, or in the coarse-grained space-time.
Based on this approach, we have worked out a suggested version of
a fractional Schrödinger equation, with a lowest-order relativistic
correction, obtained starting from a fractional wave equation \cite{Cresus-Jose-Helayel-caos}
to which a mass term has been adjoined, to give us a fractional Klein
Gordon equation (FKGE), and also with the help of the definition of
some fractional operators and Mc'Laurin expansion. By a plane wave
ansatz of solutions, we have obtained fractional versions of Bohmian
equations to describe the particle dynamics associated with Bohmian
mechanics theory and physics, in the space of non-differentiable solution
functions to the referred fractional Schrödinger equation with the
lowest-order relativistic correction.

As pointed out by Jumarie, non-differentiability and randomness are
mutually related in their nature, in such a way that studies in fractals
on the one hand and fractional Brownian motion on the other hand are
often parallel in the same paper. A function which is continuous everywhere
but is nowhere differentiable necessarily exhibits random-like or
pseudo-random-features, in the sense that various samplings of this
functions on the same given interval will be different. This may explain
the huge amount of literature which extends the theory of stochastic
differential equation to stochastic dynamics driven by fractional
Brownian motion.

The most natural and direct way to question the classical framework
of physics is to remark that in the space of our real world, the generic
point is not infinitely small (or thin) but rather has a thickness.
A coarse-grained space is a space in which a generic point is not
infinitely thin, but rather has a thickness; and here this feature
is modelled as a space in which the generic differental is not $dx$,
but rather $(dx)^{\alpha}$and likewise for coarse grained with respect
to the time variable t. It is noteworthy, at this stage, to highlight
the interesting work by Nottale \cite{Nottale}, where the notion
of fractal space-time is first introduced.

In our work, the most important rules in the MRL definition used here
is that the derivative of constant is zero, we can use it so much
for differentiable as non differentiable functions, it has simple
chain and Leibniz rules that are similar to integer derivatives. 

Our paper is outlined as follows: In Section 2, we review the development
of the low-relativistic correction to the integer order Schrödinger
equation and discuss the fractional Klein Gordon equation. In Section
3, we develop the low-relativistic fractional Schrödinger equation
and present the fractional continuity equation. In Section 4 we work
out the fractional continuity equation. Section 5 is devoted to the
development of the fractional Bohmian equations with low-relativist
limit. Finally, in Section 6 we cast our the concluding comments and
prospects for further investigation.

\section{Lowest-Order Relativistic Corrections to the Integer Schrödinger
Equation and the Fractional Klein Gordon Equation}

If we start off from the well-known relativistic relation

\begin{equation}
E=\sqrt{\vec{p}^{2}c^{2}+m^{2}c^{4}},\label{eq:Energia Momentum}
\end{equation}

\begin{equation}
\begin{cases}
E= & m_{0}c^{2}\gamma;\\
\overrightarrow{p}= & m_{0}\vec{v}\gamma
\end{cases}\qquad\gamma=\frac{1}{\sqrt{1-\frac{v^{2}}{c^{2}}}}.
\end{equation}

We readly get that $\frac{\vec{p}c}{E}=\frac{\vec{v}}{c}.$ So, in
teh non-relativistic regime ($\vec{\left|v\right|}\ll c$), $\vec{\left|p\right|}c\ll E$
and so the following approximation can be adopted:

\begin{eqnarray}
E-mc^{2} & \equiv & \varepsilon_{nr}\cong\frac{p^{2}}{2m}-\frac{p^{4}}{8m^{3}c^{2}}=\nonumber \\
 & = & \frac{\left(pc\right)^{2}}{2mc^{2}}\left[1-\left(\frac{pc}{2mc^{2}}\right)^{2}\right],
\end{eqnarray}

where $\varepsilon_{nr}$ stands for the non-relativistic kinetic
energy. Here, it is worthy of notice that the lowest-order relativistic
limit corresponds to momenta such that other terms $\vec{\left|p\right|}c\gg2mc^{2},$
that is the threshold energy for a pair creation. The fractional approach
here is still justified by the argumentation that the particle described
by this formalism is actually a pseudo-particle that carries the information
of the media and the kind interaction implicit in the equation that
describes his evolution. This pseudo-particle is then ``dressed''
with information about media and interactions, and the solutions of
the fractional equation are, like the Green functions in condensed
matter physics, carrying additional information about iterations and
media. Then, even if the media is not fractal, due to not so high
energy regime, the fractional approach still makes sense to describe
the evolutions of a pseudo-particle. This means that essentially there
is not an isolated particle in the fractional approach context but
an pseudo-particle ``dressed'' with information about the fields
and interactions in the media.

Now defining the quantum mechanics one dimensional operators, energy
and linear momentum, as usual

\begin{equation}
\begin{cases}
\widehat{E}=i\hbar\frac{\partial}{\partial t}\\
\widehat{p}=-i\hbar\frac{\partial}{\partial x}
\end{cases};
\end{equation}
 we will obtain the Schrödinger with lowest-order relativistic correction
that reads,

\begin{equation}
i\hbar\frac{\partial}{\partial t}\psi\left(x,t\right)=-\left(\frac{\hbar^{2}}{2m}\right)\frac{\partial^{2}}{\partial x^{2}}\psi\left(x,t\right)-\left(\frac{\hbar^{4}}{8m^{3}c^{2}}\right)\frac{\partial^{4}}{\partial x^{4}}\psi\left(x,t\right)+V\psi\left(x,t\right).\label{eq:Schrodinger Inteira Corrigida}
\end{equation}

With the lowest-order relativistic correction to Schrödinger equation,
we then construct in the sequence the fractional Klein Gordon equation
by adjoined a mass term to the fractional wave equation.

\subsection*{The Fractional Klein Gordon Equation }

In a recent paper \cite{Cresus-Jose-Helayel-caos}, we have obtained
in a natural way the fractional wave equation.

Now, we shall write down a fractional version of the Klein Gordon
equation by the addition of the mass term to the fractional wave equation,
considering adequate dimension scale factors, in order to gain some
insight to about the fractional quantum operator to be used.

The usual KG equation reads

\begin{equation}
\frac{1}{c^{2}}\frac{\partial^{2}}{\partial t^{2}}\psi\left(x,t\right)-\frac{\partial^{2}}{\partial x^{2}}\psi\left(x,t\right)+\frac{m^{2}c^{2}}{\hbar^{2}}\psi\left(x,t\right)=0.
\end{equation}

Fractional Klein-Gordon equation \cite{Casimir-KG-LIM} and fractional
Dirac equation have been studied by several authors over the past
decade \cite{Plyushchay-Cubic root of Klein-Gordon,Raspini,Zavada}.
Some articles have been dealing with fractional power of D\textasciiacute{}Alembertian
operator used in the non local kinetic terms Lagrangian field theory
in the (2+1)-dimensional bosonization and also to study the effective
field theory, which has some degrees of freedom integrated out from
the underlying local theory \cite{Marino 1,Barci1,Dalvit}. The canonical
quantization of fractional massless and massive fields has been studied
by some authors \cite{Marino,D. Barci} and quantization of fractional
Klein-Gordon field and fractional gauge field based on Nelson\textquoteright{}s
stochastic mechanics and Parisi-Wu stochastic quantization procedure
at zero and positive temperature have been considered \cite{Lim2,Lim3}.
An axiomatic approach to fractional Klein-Gordon field, where properties
of the n-point Schwinger or Euclidean Green functions and their analytic
continuation to the corresponding n-point Wightman functions were
studied by \cite{Albeverio,Grothaus}. 

The fractional KG equation can be written here, in an similar manner
as in ref. \cite{Muslih-1}, but with different fractional orders
in space and time, as 

\begin{equation}
\frac{1}{c^{2\beta}}\frac{\partial^{2\beta}}{\partial t^{2\beta}}\psi\left(x,t\right)-M_{x,\alpha}^{2}\frac{\partial^{2\alpha}}{\partial x^{2\alpha}}\psi\left(x,t\right)+\frac{m^{2\beta}c^{2\beta}}{\hbar^{2\beta}}\psi\left(x,t\right)=0.\label{eq:Fractional KG 1}
\end{equation}

The diffusion factor $M_{x,\alpha}$ is here introduced for dimensional
consistency reasons. This equation has also to be consistent with
an fractional relativistic energy-momentum equation, given by

\begin{equation}
E_{\beta}=\sqrt{\vec{p}^{2\alpha}c^{2\alpha}+m^{2\beta}c^{4\beta}}.\label{eq:Energia Momentum Fracional}
\end{equation}

Now, with these considerations, we shall expand the momentum energy
of eq.\eqref{eq:Energia Momentum Fracional}in terms of an integer
Mc'Laurin series and, after the substitution of fractional quantum
operators, obtain the fractional Schrödinger equation with lowest-order
relativistic correction term.

\section{Fractional Schrödinger Equation with Lowest-Order Relativistic Correction}

A method first used for the attainement of a fractional Scrodinger
equation was the path integral over the Lèvy paths formalisms \cite{Levy path-Laskin,Laskin2-Fractional Schrodinger equation}
where a fractional generalization of the Schrödinger equation in terms
of the quantum Riesz fractional derivative was obtained and there
have been analyzed the energy spectra of a hydrogen-like atom and
of a fractional oscillator in the semi-classical approximation and
the parity conservation law. The argumentation to achieve equation
\cite{Argumentation L=0000E8vy} was that a the path integral over
Brownian trajectories leads to the well-known Schrödinger equation,
then the path integral over Lèvy trajectories leads to the space fractional
Schrödinger equation. Other versions of Schrödinger equation were
obtained \cite{Naber} considering only a time fractional Schrödinger
equation in the sense of a Caputo fractional time derivative formalism.
A version of generalized fractional Schrödinger, with space-time fractional
derivatives in the sense of Caputo and Riesz fractional derivatives,
was studied in ref. \cite{Generalized Schrodinger eq} and solved
for free particle and square well potential with integral transform
methods. The fractional Schrödinger equation can also be obtained
by methods like a fractional variational method in the context of
a Lagrangian formulation or by a fractional Klein Gordon equation\cite{Muslih-1}. 

Here we adopt the MRL approach for fractional derivatives that is
less restrictive than other definitions, to obtain the lowest-order
relativistic correction to a fractional Schrödinger equation, with
different orders for the fractional derivatives in time and space,
by means of a fractional Klein Gordon equation that, by other way,
was obtained from a fractional wave equation in our recent work \cite{Cresus-Jose-Helayel-caos}.

The main rules used here in the MRL approach are summarized as

\paragraph*{\textmd{$D^{\alpha}K=0,$ $K$ is constant, $D^{\alpha}x^{\gamma}=\frac{\Gamma(\gamma+1)}{\Gamma(\gamma+1-\alpha)}x^{\gamma-\alpha},\;\gamma>0,$
derivative of power function,$(u(x)v(x))^{(\alpha)}=u^{(\alpha)}(x)v(x)+u(x)v^{(\alpha)}(x)$
is the Leibniz rule. The chain rule for non differentiable functions
is written as}}

\begin{equation}
\frac{d^{\alpha}}{dx^{\alpha}}f[u(x)]=\frac{d^{\alpha}}{du^{\alpha}}f\,\left(\frac{d}{dx}u\right)^{\alpha},\label{eq:Chainrule nondif func-1}
\end{equation}
where $f$ is \textgreek{a}-differentiable and $u$ is differentiable
with respect to $x$ and, for coarse-grained space-time as

\begin{equation}
\frac{d^{\alpha}}{dx^{\alpha}}f[u(x)]=\frac{d}{du}f\,\frac{d^{\alpha}}{dx^{\alpha}}u,\label{eq:chain rule space-time coarse-1}
\end{equation}
where $f(u(x))$is not differentiable w.r.t $x$ but it is differentiable
w.r.t $u$, and $u$is not differentiable w.r.t $x$.

For further details, the readers can follow the refs. \cite{Livro Jumarie,Jumarie- Acta Sin}
which contain all the basic for the formulation of a fractional differential
geometry in coarse-grained space, and refers to an extensive use of
coarse-grained phenomenon.

Its is worthy to point out that the Leibniz rule used here is a good
approximation that cames from the first two terms of the fractional
Taylor series development, that holds only for nondifferentiable functions
\cite{Jumarie- Acta Sin} and are as good and approximated as the
classical integer one. Here, a comment is pertinent: the fractional
MRL approach for non-differentiable functions has similar rules and
has definition with a mathematical limit operation comparable to certain
definitions of local fractional derivatives, as that introduced by
Kolwankar and Gandal \cite{Kolwankar1,Kolwankar 2,Kolwankar 3} with
some studies in the literature. For example, the works of Refs. \cite{calculus of local fractional derivatives,On the local fractional derivative,Carpintieri 1}
or the approachs with Hausdorff derivative, also called fractal derivative
\cite{Hausdorff or fractal derivative,comparative-Fractal -Fractional},
that can be applied to power-law phenomena and the recently developed
$\alpha-derivative$ \cite{Kobelev}. The MRL approach seems to us
to be an integral version of the calculus mentioned above and all
of them deserve to be more deeply investigated, under a mathematical
point of view, in order to give exact differences and similarities
respect to the traditional fractional calculus with Riemann-Liouville
or Caputo definition and with local fractional calculus and even fractional
q-calculus \cite{Metzler,Richard Herrmann,Caceres 2004,Grigolini-1999},
as well as in the comparative point of view of physics\cite{comparative-Fractal -Fractional,Caceres 2004,Grigolini-1999,West-Aging,Tsallis},
for the scope of applicablities.

We think that the referred alternative formalisms can be used to the
attainment of results similar or with similarities to some of those
here obtained \cite{fractional quantum potential} and this is a good
indication that our results are more general and not only dependent
and provided by an specific formalism.

In this work, we construct the fractional Schrödinger equation based
on operator proposed in view of the fractional Klein Gordon equation.

Developing the eq.\eqref{eq:Energia Momentum Fracional} in McLaurin's
series, doing $f(x)=(1+x_{\alpha,\beta})^{1/2}$ and assuming that
$\, f^{(\alpha k)}(x)$ have sequential character like 
\begin{equation}
f^{(2\alpha)}(x)=\frac{\partial^{2\alpha}}{\partial x^{2\alpha}}=\frac{\partial^{\alpha}}{\partial x^{\alpha}}\frac{\partial^{\alpha}}{\partial x^{\alpha}},
\end{equation}

\begin{equation}
f(x)=(1+x_{\alpha,\beta})^{1/2}\cong1+\frac{1}{2}x_{\alpha,\beta}-\frac{1}{8}x_{\alpha,\beta}^{2}+\mathcal{O}(x_{\alpha,\beta}^{3}).
\end{equation}

Since the semi group properties for fractional derivatives does not
hold in general , we used the Miller-Ross sequential derivative \cite{Miller-Ross}
in the MRL sense. Incidentally, the Miller-Ross sequential derivative
is a systematic procedure that carries out a fractional higher-order
derivative while avoiding the recursive application of many single
derivatives taken after each other. Moreover, we took the option to
carry out the sequence of derivatives in the cascade form, in MRL
sense, as done in the work of ref. \cite{Jumarie2,Jumarie- Acta Sin}. 

Here, we propose the operators:

\begin{equation}
\begin{cases}
\widehat{E}_{\beta}=i\left(\hbar\right)^{\beta}\frac{\partial^{\beta}}{\partial t^{\beta}}\\
\widehat{p}_{\alpha}=-i\left(\hbar\right)^{\alpha}M_{x,\alpha}\frac{\partial^{\alpha}}{\partial x^{\alpha}}
\end{cases};\label{eq:Fractional Energy-Momenta Operators}
\end{equation}

It can be verified that the fractional quantum operator proposed above,
when substituted into the equation eq.\eqref{eq:Energia Momentum Fracional}
will give the KG equation eq.\eqref{eq:Fractional KG 1}.

Note that $M_{x,\alpha}$ factor becomes dimensionless when $\alpha$
equal to $1$ and its mass dimension is in general $\alpha-$dependent.

Now, evidencing the term $m^{\beta}c^{2\beta}$in the eq.\eqref{eq:Energia Momentum Fracional}
and expanding in therms of McLaurin's series in $x_{\alpha,\beta}=\frac{p^{2\alpha}c^{2\alpha}}{m^{2\beta}c^{4\beta}},$
by substitution of the fractional operators in eq. \eqref{eq:Fractional Energy-Momenta Operators},
we are lead to one possible representation of the fractional Schrödinger
equation given by 
\begin{multline}
\mbox{\ensuremath{\begin{aligned}i\left(\hbar\right)^{\beta}\frac{\partial^{\beta}}{\partial t^{\beta}}\psi\left(x,t\right) & =-M_{x,\alpha}^{2}\frac{\hbar^{2\alpha}}{2m^{\beta}}\frac{c^{2\alpha}}{c^{2\beta}}\frac{\partial^{2\alpha}}{\partial x^{2\alpha}}\psi\left(x,t\right)+V_{\alpha,\beta}\psi\left(x,t\right)+\\
 & -\frac{1}{8}M_{x,\alpha}^{4}\frac{\hbar^{4\alpha}}{m^{3\beta}}\frac{c^{4\alpha}}{c^{6\beta}}\frac{\partial^{4\alpha}}{\partial x^{4\alpha}}\psi\left(x,t\right)
\end{aligned}
}}\label{eq:Schrodinger Fractional-corrigida}
\end{multline}
where the notation is assumed $\frac{\partial^{4\alpha}}{\partial x^{4\alpha}}=\frac{\partial^{\alpha}}{\partial x^{\alpha}}\frac{\partial^{\alpha}}{\partial x^{\alpha}}\frac{\partial^{\alpha}}{\partial x^{\alpha}}\frac{\partial^{\alpha}}{\partial x^{\alpha}},$
$\frac{\partial^{2\alpha}}{\partial x^{2\alpha}}=\frac{\partial^{\alpha}}{\partial x^{\alpha}}\frac{\partial^{\alpha}}{\partial x^{\alpha}}$,
since the semi-group properties for additive in the orders of the
derivatives may not hold, as previously commented.

\section{Fractional Continuity Equation}

It is well-known that in standard quantum mechanics continuity equation
has a very important figure, it represents a conservation law. In
the context of standard quantum mechanics, in the Copenhagen he, we
are lead to the conservation of the probability density. But, in the
context of a fractional quantum mechanics, the meaning of a fractional
continuity equation is not quite clear and require some analysis.
Since we are in the interaction picture and are handling with pseudo-particles
or ''dressed'' particles, the fractional continuity equation could
give us in true, the revelation that it exist a dissipation implicit
in the fractional evolution equations, specially if the orders of
derivatives in space and time were different from each other. This
could mean that the fractional equations can be thought of related
to some effective theories. The known and unknown information about
interactions and the media could be acconted for in fractionality.
When the integer order limit for derivatives are reached, the conservation
law emerges, the dissipation are no more present in the theory and
certain symmetries could be reestablished.

We expect that future scientific investigations may clarify more the
real meaning of the fractional continuity equation.

To obtain our fractional continuity equation we now proceed as follows:
the conjugate of fractional Schrödinger equation reads

\begin{multline}
\mbox{\ensuremath{\begin{aligned}-i\left(\hbar\right)^{\beta}\frac{\partial^{\beta}}{\partial t^{\beta}}\psi\left(x,t\right) & =-M_{x,\alpha}^{2}\frac{\hbar^{2\alpha}}{2m^{\beta}}\frac{c^{2\alpha}}{c^{2\beta}}\frac{\partial^{2\alpha}}{\partial x^{2\alpha}}\psi\left(x,t\right)+V_{\alpha,\beta}\psi\left(x,t\right)+\\
 & -\frac{1}{8}M_{x,\alpha}^{4}\frac{\hbar^{4\alpha}}{m^{3\beta}}\frac{c^{4\alpha}}{c^{6\beta}}\frac{\partial^{4\alpha}}{\partial x^{4\alpha}}\psi\left(x,t\right)
\end{aligned}
}}\label{eq:Schrodinger Fractional-corrigida-Conjugada}
\end{multline}

Defining the probability as $P=\psi^{*}\left(x,t\right)\psi\left(x,t\right).$

Multiplying \eqref{eq:Schrodinger Fractional-corrigida} by $\psi^{*}\left(x,t\right)$
and equation \eqref{eq:Schrodinger Fractional-corrigida-Conjugada}
by $-\psi\left(x,t\right)$, after adding both equations, we obtain

\begin{multline}
\mbox{\ensuremath{\begin{aligned}i\left(\hbar\right)^{\beta}\frac{\partial^{\beta}}{\partial t^{\beta}}\psi^{*}\left(x,t\right)\psi\left(x,t\right) & =-M_{x,\alpha}^{2}\frac{\hbar^{2\alpha}}{2m^{\beta}}\frac{c^{2\alpha}}{c^{2\beta}}\left[\psi^{*}\left(x,t\right)\frac{\partial^{2\alpha}}{\partial x^{2\alpha}}\psi\left(x,t\right)-\psi\left(x,t\right)\frac{\partial^{2\alpha}}{\partial x^{2\alpha}}\psi^{*}\left(x,t\right)\right]+\\
 & -\frac{1}{8}M_{x,\alpha}^{4}\frac{\hbar^{4\alpha}}{m^{3\beta}}\frac{c^{4\alpha}}{c^{6\beta}}\left[\psi^{*}\left(x,t\right)\frac{\partial^{4\alpha}}{\partial x^{4\alpha}}\psi\left(x,t\right)-\psi\left(x,t\right)\frac{\partial^{4\alpha}}{\partial x^{4\alpha}}\psi^{*}\left(x,t\right)\right]
\end{aligned}
}}.\label{eq:Schrodinger Fractional-corrigida adicionada com Conjugada}
\end{multline}

After some algebra, the latter equation can be written as

\begin{equation}
\mbox{\ensuremath{\begin{aligned}\frac{\partial^{\beta}}{\partial t^{\beta}}\rho\left(x,t\right)+\frac{\partial^{\alpha}}{\partial x^{\alpha}}J(x,t) & =0,\\
\\
\end{aligned}
}}\label{eq:Continuidade Fracional}
\end{equation}
where $\rho\left(x,t\right)\equiv\psi^{*}\left(x,t\right)\psi\left(x,t\right),$
and

\begin{multline}
\mbox{\ensuremath{\begin{aligned}J= & M_{x,\alpha}^{2}\frac{\hbar^{2\alpha}}{2m^{\beta}i\hbar^{\beta}}\frac{c^{2\alpha}}{c^{2\beta}}J'+\\
 & +\frac{1}{8}M_{x,\alpha}^{4}\frac{\hbar^{4\alpha}}{m^{3\beta}i\hbar^{\beta}}\frac{c^{4\alpha}}{c^{6\beta}}\frac{\partial^{\alpha}}{\partial x^{\alpha}}\left\{ \left[J'-2\left(\frac{\partial^{\alpha}\psi^{*}\left(x,t\right)}{\partial x^{\alpha}}\frac{\partial^{\alpha}\psi\left(x,t\right)}{\partial x^{\alpha}}\right)\right]+4\left(\frac{\partial^{\alpha}\psi\left(x,t\right)}{\partial x^{\alpha}}\frac{\partial^{2\alpha}\psi^{*}\left(x,t\right)}{\partial x^{2\alpha}}\right)\right\} ,
\end{aligned}
}}\label{eq:Schrodinger Fractional-corrigida adicionada com Conjugada-2}
\end{multline}

with $J'\equiv\left[\psi^{*}\left(x,t\right)\frac{\partial^{\alpha}}{\partial x^{\alpha}}\psi\left(x,t\right)-\psi\left(x,t\right)\frac{\partial^{\alpha}}{\partial x^{\alpha}}\psi^{*}\left(x,t\right)\right]$.

The equation \eqref{eq:Continuidade Fracional} shows that the probability
is conserved in the fractional sense.

Taking $\alpha=\beta=1$, we obtain the integer continuity equation
with the lowest-order relativistic correction.

\section{Fractional Quantum Potential with Lowest-Order Relativistic Correction
terms}

Now, we shall build up the fractional Bohmian equations, by parametrizing
the solution of eq.\eqref{eq:Schrodinger Fractional-corrigida} as
below:
\begin{equation}
\Psi({\bf r},t)=R({\bf r},t)e^{iS({\bf r},t)/\hbar},\label{e2}
\end{equation}
where $R$ and $S$ are the amplitude of probability density and phase
of $\Psi$, respectively, both being real-valued functions. Substituting
this relation into the fractional Schrödinger\textquoteright{}s equation
and multiplying by $e^{-iS({\bf r},t)/\hbar}$, after some algebra
and taking real and imaginary parts, we get two equations that lead
to a fractional version of Bohmian Mechanics, including the its lowest-order
relativistic correction limit.

Now, proceeding as described above, two equations are obtained:

a) for the real part:

\begin{multline}
-M_{x,\alpha}^{2}\frac{\hbar^{2\alpha}}{2m^{\beta}}\frac{c^{2\alpha}}{c^{2\beta}}\frac{1}{R\left(x,t\right)}\frac{\partial^{\alpha}}{\partial x^{\alpha}}\frac{\partial^{\alpha}}{\partial x^{\alpha}}R\left(x,t\right)+M_{x,\alpha}^{2}\frac{\hbar^{2\alpha}}{2m^{\beta}}\frac{c^{2\alpha}}{c^{2\beta}}\frac{1}{\hbar^{2}}\left(\frac{\partial^{\alpha}S}{\partial x^{\alpha}}\right)^{2}+\hbar^{\beta-1}\frac{\partial^{\beta}}{\partial t^{\beta}}S\left(x,t\right)+V+\\
\\
-\frac{1}{8}M_{x,\alpha}^{4}\frac{\hbar^{4\alpha}}{m^{3\beta}}\frac{c^{4\alpha}}{c^{6\beta}}\frac{1}{R}\left[\frac{1}{\hbar^{4}}R(S^{(\alpha)})^{4}+R^{(4\alpha)}-\frac{4}{\hbar^{2}}RS^{(\alpha)}S^{(3\alpha)}-\frac{12}{\hbar^{2}}R^{(\alpha)}S^{(\alpha)}S^{(2\alpha)}+-\frac{3}{\hbar^{2}}R{}^{2}(S^{(2\alpha)})^{2}-\frac{6}{\hbar^{2}}R^{(2\alpha)}(S^{(\alpha)})^{2}\right]=0,\label{eq:firstBohmEq}
\end{multline}

\pagebreak{}

b) for the imaginary part:

\begin{align}
\frac{\partial^{\beta}R^{2}}{\partial t^{\beta}}+2M_{x,\alpha}^{2}\frac{1}{2m^{\beta}}\frac{c^{2\alpha}}{c^{2\beta}}\hbar^{2\alpha-\beta-1}\frac{\partial^{\alpha}}{\partial x^{\alpha}}\left(R^{2}\frac{\partial^{\alpha}S}{\partial x^{\alpha}}\right)+\nonumber \\
-\frac{1}{8}M_{x,\alpha}^{4}\frac{\hbar^{4\alpha}}{m^{3\beta}}\frac{c^{4\alpha}}{c^{6\beta}}\frac{i}{\hbar^{4}}\left(\frac{-2R}{\hbar^{\beta}}\right)\left[R\hbar^{3}S^{(4\alpha)}-4\hbar R^{(\alpha)}(S^{(\alpha)})^{3}+4\hbar^{3}R^{(\alpha)}S^{(3\alpha)}+6\hbar^{3}R^{(2\alpha)}S^{(2\alpha)}\right] & =0.\label{eq:SecBohm Eq}
\end{align}

The first term in the left-hand side of eq.\eqref{eq:firstBohmEq}
can be called fractional quantum potential by the presence of Planck
constant and fractional derivatives

\begin{multline}
Q_{\alpha}(x,t)\equiv-M_{x,\alpha}^{2}\frac{\hbar^{2\alpha}}{2m^{\beta}}\frac{c^{2\alpha}}{c^{2\beta}}\frac{1}{R\left(x,t\right)}\frac{\partial^{\alpha}}{\partial x^{\alpha}}\frac{\partial^{\alpha}}{\partial x^{\alpha}}R\left(x,t\right)+\\
\\
-\frac{1}{8}M_{x,\alpha}^{4}\frac{\hbar^{4\alpha}}{m^{3\beta}}\frac{c^{4\alpha}}{c^{6\beta}}\frac{1}{R}\left[R^{(4\alpha)}-\frac{4}{\hbar^{2}}RS^{(\alpha)}S^{(3\alpha)}-\frac{12}{\hbar^{2}}R^{(\alpha)}S^{(\alpha)}S^{(2\alpha)}+-\frac{3}{\hbar^{2}}R{}^{2}(S^{(2\alpha)})^{2}-\frac{6}{\hbar^{2}}R^{(2\alpha)}(S^{(\alpha)})^{2}\right]\label{eq:FracQuantPot-1}
\end{multline}

With this definition, the eq. \eqref{eq:firstBohmEq}can be rewritten
as

\begin{equation}
Q_{\alpha}(x,t)+V+M_{x,\alpha}^{2}\frac{\hbar^{2\alpha}}{2m^{\beta}}\frac{c^{2\alpha}}{c^{2\beta}}\frac{1}{\hbar^{2}}\left(\frac{\partial^{\alpha}S}{\partial x^{\alpha}}\right)^{2}+-\frac{1}{8}M_{x,\alpha}^{4}\frac{\hbar^{4\alpha}}{m^{3\beta}}\frac{c^{4\alpha}}{c^{6\beta}}\frac{1}{\hbar^{4}}(S^{(\alpha)})^{4}=-\hbar^{\beta-1}\frac{\partial^{\beta}}{\partial t^{\beta}}S\left(x,t\right),
\end{equation}
deriving this equation with respect to $x^{\alpha}$, interchanging
spatial and time ordering of derivatives and considering both fractional
derivative orders equals, that is, $\lyxmathsym{\textgreek{a}}=\lyxmathsym{\textgreek{b}}$,
we obtain

\begin{equation}
-\frac{\partial^{\alpha}}{\partial x^{\alpha}}\left(Q_{\alpha}(x,t)+V\right)=\frac{\partial^{\alpha}}{\partial x^{\alpha}}\left[M_{x,\alpha}^{2}\frac{\hbar^{2\alpha}}{2m^{\beta}}\frac{c^{2\alpha}}{c^{2\beta}}\frac{1}{\hbar^{2}}\left(\frac{\partial^{\alpha}S}{\partial x^{\alpha}}\right)^{2}-\frac{1}{8}M_{x,\alpha}^{4}\frac{\hbar^{4\alpha}}{m^{3\beta}}\frac{c^{4\alpha}}{c^{6\beta}}\frac{1}{\hbar^{4}}(S^{(\alpha)})^{4}\right]+\hbar^{\alpha-1}\frac{\partial^{\alpha}}{\partial t^{\alpha}}\frac{\partial^{\alpha}}{\partial x^{\alpha}}S\left(x,t\right).
\end{equation}

Defining the fractional moment as

\begin{equation}
\mathit{\mathit{p}_{\alpha}}=\hbar^{\alpha-1}\frac{\partial^{\alpha}S}{\partial x^{\alpha}},
\end{equation}
noting that in the lowest order in $\alpha$

\begin{equation}
\frac{d^{\alpha}\mathit{p_{\alpha}}}{dt^{\alpha}}=\frac{\partial^{\alpha}\mathit{p}_{\alpha}}{\partial x^{\alpha}}\left(\frac{dx}{dt}\right)^{\alpha}+\frac{\partial^{\alpha}\mathit{p}_{\alpha}}{\partial t^{\alpha}},
\end{equation}
with a similar the definition of the fractional velocity \cite{Jumarie-Lagrang Fract},
that relates it to a fractional linear momentum,

\begin{equation}
v_{\alpha}=\left(\frac{dx}{dt}\right)^{\alpha}=\lambda_{\alpha,\beta}\mathit{p}_{\alpha},
\end{equation}
with $\lambda_{\alpha,\beta}=\left(M_{x,\alpha}\frac{c^{\alpha}}{c^{\beta}}\right)^{-1}$we
will have that

\begin{equation}
-\frac{\partial^{\alpha}}{\partial x^{\alpha}}\left(Q_{\alpha}(x,t)+V\right)\equiv F_{\alpha}.
\end{equation}
where $F_{\alpha}$ is defined as the fractional force. The equation
above gives us a Newtonian-like fractional dynamical equation, that
coincide with $\frac{d^{\alpha}\mathit{p_{\alpha}}}{dt^{\alpha}}$
if $\alpha=1$ and we do not consider the higher order term.

Defining also the fractional mechanical energy and the kinetic energy,
respectively as

\begin{equation}
E_{\alpha}(x,t)=-\hbar^{\alpha-1}\frac{\partial^{\alpha}}{\partial t^{\alpha}}S\left(x,t\right),
\end{equation}
and

\begin{equation}
K_{\alpha}(x,t)=M_{x,\alpha}^{2}\frac{\hbar^{2\alpha}}{2m^{\beta}}\frac{c^{2\alpha}}{c^{2\beta}}\frac{1}{\hbar^{2}}\left(\frac{\partial^{\alpha}S}{\partial x^{\alpha}}\right)^{2}-\frac{1}{8}M_{x,\alpha}^{4}\frac{\hbar^{4\alpha}}{m^{3\beta}}\frac{c^{4\alpha}}{c^{6\beta}}\frac{1}{\hbar^{4}}(S^{(\alpha)})^{4}.\label{eq:Kinectic term}
\end{equation}

In terms of these and the quantum potential, we can rewrite eq. \eqref{eq:firstBohmEq}
as

\begin{equation}
E_{\alpha}(x,t)=K_{\alpha}(x,t)+Q_{\alpha}(x,t)+V.
\end{equation}

It is important to note that if we make \textgreek{a}=1, all the results
are in complete accord with standard Bohmian mechanics theory with
the inclusion of lower relativistic correction terms.

Another point to highlight concerns energy conservation. If we assumee
for the phase S a dependence like a power of time,

\begin{equation}
S\left(x,t\right)=E\cdot\hbar(f-t^{\alpha}),
\end{equation}
where $E$ is a multiplicative constant and $f$ is some functions
depending explicitly only on $x$, then we obtain for the fractional
energy

\begin{equation}
E_{\alpha}(x,t)=-\hbar^{\alpha-1}\frac{\partial^{\alpha}}{\partial t^{\alpha}}S\left(x,t\right)=E\hbar^{\alpha}\Gamma(\alpha+1),
\end{equation}
 that is a constant. The the fractional energy can be conserved by
an appropriate choice of phase.

\subsection*{De Broglie relations}

If we write for the phase S a dependence like a power of time,

\begin{equation}
S\left(x,t\right)=(kx)^{\alpha}\pm(\omega t)^{\alpha},
\end{equation}
 we will have for the energy 

\begin{equation}
E_{\alpha}(x,t)=\hbar^{\alpha}\Gamma(\alpha+1)(\omega)^{\alpha}
\end{equation}

Note that when $\alpha=1,$ then $E=\hbar\omega.$

Inserting these phase S into the eq. \eqref{eq:Kinectic term} leads
to

\begin{eqnarray}
K_{\alpha}(x,t) & = & M_{x,\alpha}^{2}\frac{\hbar^{2\alpha}}{2m^{\alpha}}\frac{1}{\hbar^{2}}\left(\Gamma(\alpha+1)k^{\alpha}\right)^{2}-\frac{1}{8}M_{x,\alpha}^{4}\frac{\hbar^{4\alpha}}{m^{3\alpha}}\frac{1}{c^{2\alpha}}\frac{1}{\hbar^{4}}(\Gamma(\alpha+1)k^{\alpha})^{4}=\nonumber \\
 & = & M_{x,\alpha}^{2}\frac{1}{2m^{\alpha}}\frac{1}{\hbar^{2}}\left[\Gamma(\alpha+1)\hbar^{\alpha}k^{\alpha}\right]^{2}+\mathcal{O\left(\mbox{\ensuremath{p^{4}}}\right)}.
\end{eqnarray}

Defining 

\begin{eqnarray}
p_{\alpha} & = & M_{x,\alpha}\Gamma(\alpha+1)\hbar^{\alpha}k^{\alpha},
\end{eqnarray}

which reduces to de Broglie relations of ordinary quantum mechanics
whenever $\alpha=1.$

\section{Concluding Comments}

There has been considerable interest over the past recent years in
the so-called theory of \textquotedbl{}weak\textquotedbl{} quantum
measurements, whose aim seems to be to measure the average value of
a quantum observable while negligibly disturbing the measured system
\cite{Aharavov1,Aharavov2,Hosten,Lundeen,Yokota}. Very recently,
experimental observation of trajectories of a photon in a double-slit
interferometer was reported, which displayed the qualitative features
predicted in the de Broglie-Bohm interpretation \cite{SachaKoc1,SachaKoc2}.

Possibilities like connections with a quantum gravity theory emerges
from the fact that an modified fractional Newtonian equation could
be connected with a fractional Newtonian dynamics similar to MOND
of Mordehai Milgrom \cite{MOND}. The natural emergence of a fractional
Newtonian equation implicitly involves a non-local theory leading
to a Newtonian law with memory, a characteristic of fractional derivatives.
Also, the fractional energy reinforces the expectation of the presence
of quantum effects. Those effects can be also associated with collective
behavior in a fractal space-time tissue, where fluctuations can give
rise to excitations like tisons or fractons.

Also, a version of fractional de Broglie relations naturally comes
out from our equations and we recover the integer relations in the
convenient limit. In connection with the probability conservation,
in the fractional case, we have worked out, to the lowest order in
the relativistic correction, the fractional probability current. The
probability can be conserved in this non-differentiable space-time
if we consider a fractional version of continuity equation that reduces
to the standard one in the integer limit or, in other words, integer
dimensions. As an outlook for a forthcoming work, solutions with the
Mittag-Lefller instead of exponential solutions, shall be analyzed
in two possibilities: non-differentiable space of solutions and coarse-grained
space-time in the argument of refereed special solution function.

\medskip{}

The professor (one of the authors), J. A. Helayël-Neto, would like
to express their gratitude to the Brazilian FAPERJ and CNPq for partial
financial support.


\begin{thebibliography}{References}
\bibitem{Richard Herrmann}Richard Herrmann, ''Common aspects of q-deformed
Lie algebras and fractional calculus'', Physica A 389 (2010) 46134622;
arXiv:1007.1084v1 {[}physics.gen-ph{]}.

\bibitem{Cresus e Everton}E.M.C Abreu and C.F.L. Godinho, Phys. Rev.
E84 026608 (2011).

\bibitem{Cresus-Jose-Helayel-caos}Cresus F.L. Godinho, J. Weberszpil,
J.A. Helayël-Neto, ''Extending the D'Alembert Solution to Space-Time
Modied Riemann-Liouville Fractional Wave Equations, Chaos, Solitons
\& Fractal, Chaos, Solitons \& Fractals, \textbf{45} 765\textendash{}771
(2012).

\bibitem{Q. Su}Q. Su, B. A. Smetanko 1 , and R. Grobe, ''Wave Packet
Motion in Relativistic Electric Fields'', Laser Physics, Vol. 8, No.
1, pp. 93\textendash{}101 (1998).

\bibitem{Dyall}K.G. Dyall, ''Inclusion of Relativistic Effects in
Electronic Structure Calculation'', Aust. J. Phys.,\textbf{39},667-78,
(1986).

\bibitem{Reither}Markus Reiher and Bernd Heß, ''Relativistic Electronic-Structure
Calculations for Atoms and Molecules'', Modern Methods and Algorithms
of Quantum Chemistry, Proceedings, Second Edition, J. Grotendorst
(Ed.), John von Neumann Institute for Computing, J\textasciidieresis{}
ulich, NIC Series, Vol. 3, ISBN 3-00-005834-6, pp. 479-505, 2000.

\bibitem{Nickisch}Nickisch L J, Durand L and Durand B, Salpeter equation
in position space: numerical solution for arbitrary confi{}ning potentials,
Phys. Rev D30(1984) 660-70; 

\bibitem{Frohlich J and Lenzmann E}Frohlich J and Lenzmann E, Blowup
for nonlinear wave equations describing boson stars, Com. Pure Appl.Math.
(2007) 0001-15;

\bibitem{Jumarie1}G. Jumarie, J. Appl. Math. \& Computing Vol. 24,
No. 1 - 2, pp. 31 - 48 (2007); Applied Mathematics Letters 22, 378385
(2009);.Guy Jumarie, \textquotedbl{}Table of some basic fractional
calculus formulae derived from a modified RiemannLiouville derivative
for non-differentiable functions\textquotedbl{}, Applied Mathematics
Letters \textbf{22} 378385 (2009).

\bibitem{Jumarie-Lagrang Fract}Guy Jumarie, \textquotedbl{}From Lagrangian
mechanics fractal in space to space fractal Schrödinger\textquoteright{}s
equation via fractional Taylor\textquoteright{}s series\textquotedbl{},
Chaos, Solitons and Fractals \textbf{41} 1590\textendash{}1604 (2009).

\bibitem{Livro Jumarie}Jumarie, G.; \textquotedbl{}White noise calculus,
stochastic calculus, coarse-graining and fractal geodesic. A unied
approach via fractional calculus and Maruyamas notation\textquotedbl{},
In Brownian motion: Theory, Modelling and Applica- tion, R.C. Earnshaw
and E.M. Riley Edit. pp 1-69, Nova Publishing, 2011.

\bibitem{Jumarie- Acta Sin}Jumarie, G. Acta Mathematica Sinica,\textbf{
}Published online: February 13, 2012; DOI: 10.1007/s10114-012-0507-3.

\bibitem{Jumarie2}Guy Jumarie, Computers and Mathematics with Applications
59, 1142-1164 (2010).

\bibitem{Nottale}Nottale L. Fractal space-time and micro physics.
Singapore: World Scientifi{}c; 1993; Nottale L. Chaos Solitons Fract
1994;4:361\textendash{}88.

\bibitem{Casimir-KG-LIM}S. C. Lim, L. P. Teo, ''Casimir Effect Associated
with Fractional Klein-Gordon Field'', arXiv:1103.1673v1 {[}math-ph{]},9
Mar 2011.

\bibitem{Plyushchay-Cubic root of Klein-Gordon}M. S. Plyushchay and
M. R. de Traubenber, Cubic root of Klein-Gordon equation, Phys. Lett.
B 477, 276\textendash{}284 (2000).

\bibitem{Raspini}A. Raspini, Simple solution of fractional Dirac
equation of order 2/3, Phys. Scr. 64, 20\textendash{}22 (2001).

\bibitem{Zavada}P. Zavada, Relativistic wave equations with fractional
derivatives and pseudodifferential operators, J. Appl. Math. 2, 163\textendash{}197
(2002).

\bibitem{Marino 1}E. C. Marino, Quantum electrodynamics of particles
on a plane and the Chern-Simons theory, Nucl. Phys. B 408, 551\textendash{}564
(1993). 

\bibitem{Barci1}G. G. Barci, C.D. Fosco and L.E. Oxman, On bosonization
in 3 dimensions, Phys. Lett. B 375, 267\textendash{}272 (1996). 

\bibitem{Dalvit}D. A. R. Dalvit and F.D. Mazzitelli, Running coupling
constants, Newtonian potential, and nonlocalities in the effective
action, Phys. Review D 50, 1001\textendash{} 1009 (1994).

\bibitem{Marino}R. L. P. G. do Amaral and E. C. Marino, Canonical
quantization of theories containing fractional powers of the DŒAlembertian
operator, J. Phys. A: Math. Gen. 25, 5183.5200 (1992). 

\bibitem{D. Barci}D. G. Barci, L. E. Oxman and M. Rocca, Canonical
quantization of non-local field equations, Int. J. Mod. Phys. A 11,
2111.2126 (1996).

\bibitem{Lim2}S. C. Lim and S. V. Muniandy, Stochastic quantization
of nonlocal fields, Phys. Lett. A 324, 396\textendash{}405 (2004).

\bibitem{Lim3}S. C. Lim, Fractional derivative quantum fields at
positive temperature, Phys- ica A 363, 269\textendash{}281 (2006).

\bibitem{Albeverio}S. Albeverio, H. Gottschalk and J-L Wu, Convoluted
generalized white noise, Schwinger functions and their analytic continuation
to Wightman functions, Rev. Math. Phys. 8, 763\textendash{}817 (1996). 

\bibitem{Grothaus}M. Grothaus and L. Streit, Construction of relativistic
quantum fields in the framework of white noise analysis, J. Math.
Phys. 40, 5387\textendash{}5405 (1999).

\bibitem{Muslih-1}Sami I. Muslih ·Om P. Agrawal · Dumitru Baleanu,''A
Fractional Schrödinger Equation and Its Solution'', Int. J. Theor.
Phys. (2010) 49: 1746\textendash{}1752

\bibitem{Levy path-Laskin}Nikolai Laskin, ''Fractional quantum mechanics
and Le\textasciiacute{}vy path integrals'', Physics Letters A 268,
298\textendash{}305 (2000).

\bibitem{Laskin2-Fractional Schrodinger equation}N. Laskin,''Fractional
Schrödinger equation'' Phys. Rev. E 66, 056108 (2002).

\bibitem{Argumentation L=0000E8vy}Xiaoyun Jiang, Haitao Qi, and Mingyu
Xu, ''Exact solutions of fractional Schrödinger-like equation with
a nonlocal term'', J. Math. Phys. 52, 042105 (2011).

\bibitem{Naber}M. Naber,''Time fractional Schrödinger equation''
J. Math. Phys. 45, 3339 (2004).

\bibitem{Generalized Schrodinger eq}Shaowei Wang and Mingyu Xu, ''Generalized
fractional Schrödinger equation with space-time fractional derivatives'',
J. Math. Phys. 48, 043502 (2007).

\bibitem{Kolwankar1}Kolwankar KM, Gangal AD. ''Local fractional calculus:
a calculus for fractal space\textendash{}time. In: Fractals: theory
and application in engineering'', Delft: Springer; 1999. 

\bibitem{Kolwankar 2} Kolwankar KM, Gangal AD. ''Fractional differentiability
of nowhere differentiable functions and dimensions'', Chaos 1996;6:505\textendash{}23.

\bibitem{Kolwankar 3} Kolwankar KM, Gangal AD., ''Local fractional
Fokker\textendash{} Planck equation'', Phys Rev Lett 1998;80:214\textendash{}7.

\bibitem{calculus of local fractional derivatives}A. Babakhani and
Varsha Daftardar-Gejji, ''On calculus of local fractional derivatives'',
J. Math. Anal. Appl. 270 (2002) 66\textendash{}79.

\bibitem{On the local fractional derivative}Yan Chen a, Ying Yanb,
Kewei Zhangc,.''On the local fractional derivative'', J. Math. Anal.
Appl. 362 (2010) 17\textendash{}33.

\bibitem{Carpintieri 1}A. Carpinteri, B. Chiaia, P. Cornetti, ''The
elastic problem for fractal media: basic theory and finite element
formulation'', Computers and Structures 82 (2004) 499\textendash{}508.

\bibitem{Hausdorff or fractal derivative}W. Chen, ''Time\textendash{}space
fabric underlying anomalous diffusion'', Chaos, Solitons and Fractals
28 (2006) 923\textendash{}929.

\bibitem{comparative-Fractal -Fractional}W. Chen, Hongguang Sun,
Xiaodi Zhang, Dean Koroak, ''Anomalous diffusion modeling by fractal
and fractional derivatives'', Computers and Mathematics with Applications
59 (2010) 17541758.

\bibitem{Kobelev}Kobelev V. arXiv: math-ph/1202.2714; Chaos, 2006,
v. 16, 043117

\bibitem{Metzler}Ralf Metzler, ''Generalized Chapman-Kolmogorov equation:
A unifying approach to the description of anomalous transport in external
fields'', Phys. Rev. E \textbf{62}, 5 (2000).

\bibitem{Caceres 2004}Adrián A. Budini and Manuel O. Cáceres, ''Functional
characterization of generalized Langevin equations'', J. Phys. A:
Math. Gen. 37 (2004) 5959\textendash{}5981

\bibitem{Grigolini-1999}Marco Buiatti, Paolo Grigolini and Anna Montagnini,
''Dynamic Approach to the Thermodynamics of Superdiffusion'', Phys.
Rev. Lett., \textbf{82},17, 3383-3387 (1999)

\bibitem{West-Aging}Gerardo Aquino, Mauro Bologna, Paolo Grigolini
and Bruce J. West, ''Aging and rejuvenation with fractional derivatives'',
Phys. Rev E \textbf{70}, 036105 (2004).

\bibitem{Tsallis}C. Tsallis, S. V. F. Levy, A. M. C. Souza, and R.
Maynard, ''Statistical-Mechanical Foundation of the Ubiquity of Levy
Distributions in Nature'', Phys. Rev. Lett. 75, 3589\textendash{}3593
(1995)

\bibitem{fractional quantum potential}Robert Caroll, ''On fractional
quantum potential'', Progress in Physics, \textbf{2}, (2012) 82-86. 

\bibitem{Miller-Ross} K.Miller and B. Ross, \textit{''An Introduction
to the Fractional Calculus and Fractional Differential Equations}'',
Willey\&Sons, INC., 1993.

\bibitem{Aharavov1} Y. Aharonov, D.Z. Albert, L. Vaidman, \textquotedbl{}How
the result of a measurement of a component of the spin of a spin-1/2
particle can turn out to be 100,\textquotedbl{} Physical Review Letters,
(1988). 

\bibitem{Aharavov2}Y. Aharonov and L. Vaidman in \textquotedbl{}Time
in Quantum Mechanics', J.G. Muga et al. eds., (Springer) 369-412 (2002)
quant-ph/0105101.

\bibitem{Hosten}O. Hosten and P. Kwiat, \textquotedbl{}Observation
of the spin Hall effect of light via weak measurements\textquotedbl{},
Science \textbf{319} 787 (2008).

\bibitem{Lundeen} J. S. Lundeen, A. M. Steinberg, \textquotedbl{}Experimental
Joint Weak Measurement on a Photon Pair as a Probe of Hardy's Paradox\textquotedbl{},
Phys. Rev. Lett. \textbf{102}, 020404 (2009).

\bibitem{Yokota}K. Yokota, T. Yamamoto, M. Koashi, N. Imoto, \textquotedbl{}Direct
observation of Hardy's paradox by joint weak measurement with an entangled
photon pair\textquotedbl{}, New J. Phys. \textbf{11}, 033011 (2009)

\bibitem{SachaKoc1}Sacha Kocsis, Sylvain Ravets, Boris Braverman,
Krister Shalm, Aephraim M. Steinberg, \textquotedbl{}Observing the
trajectories of a single photon using weak measurement\textquotedbl{},
19th Australian Instuturte of Physics (AIP) Congress, 2010

\bibitem{SachaKoc2}Sacha Kocsis, et al., \textquotedbl{}Observing
the Average Trajectories of Single Photons in a Two-Slit Interferometer\textquotedbl{},
Science \textbf{332}, 6034 1170-1173 (2011).

\bibitem{MOND}Mordehai Milgrom, ''MOND--a pedagogical review'', Acta
Phys.Polon. B32 (2001) 3613; arXiv:astro-ph/0112069v1.\end{thebibliography}
\end{document}